\documentclass[aps,prd,preprint,nofootinbib]{revtex4-1}

\usepackage{amsmath}
\usepackage{amssymb}
\usepackage{amsthm}
\usepackage{mathrsfs}
\usepackage{graphicx}
\usepackage{epstopdf}
\usepackage{fancyhdr}
\usepackage{array}
\usepackage[all]{xy}
\usepackage{eufrak}
\usepackage{euscript}
\usepackage{enumerate}
\usepackage{slashed}
\usepackage{hyperref}
\usepackage{caption}

\newcommand{\be}{\begin{equation}}
\newcommand{\ee}{\end{equation}}
\newcommand{\bea}{\begin{eqnarray}}
\newcommand{\eea}{\end{eqnarray}}

\def \v {\vec}

\begin{document}

\title{Anisotropic heavy quark potential in strongly-coupled $\mathcal{N}=4$ SYM in a magnetic field}

\author{R.~Rougemont}
\email{romulo@if.usp.br}
\affiliation{Instituto de F\'{i}sica, Universidade de S\~{a}o Paulo, S\~{a}o Paulo, SP, Brazil}

\author{R.~Critelli}
\email{renato.critelli@usp.br}
\affiliation{Instituto de F\'{i}sica, Universidade de S\~{a}o Paulo, S\~{a}o Paulo, SP, Brazil}

\author{J.~Noronha}
\email{noronha@if.usp.br}
\affiliation{Instituto de F\'{i}sica, Universidade de S\~{a}o Paulo, S\~{a}o Paulo, SP, Brazil}

\date{\today}

\begin{abstract}
In this work we use the gauge/gravity duality to study the anisotropy in the heavy quark potential in strongly coupled $\mathcal{N}=4$ Super-Yang Mills (SYM) theory (both at zero and nonzero temperature) induced by a constant and uniform magnetic field $\mathcal{B}$. At zero temperature, the inclusion of the magnetic field decreases the attractive force between heavy quarks with respect to its $\mathcal{B}=0$ value and the force associated with the parallel potential is the least attractive force. We find that the same occurs at nonzero temperature and, thus, at least in the case of strongly coupled $\mathcal{N}=4$ SYM, the presence of a magnetic field generally weakens the interaction between heavy quarks in the plasma.
\end{abstract}

\maketitle


\section{Introduction}
\label{introduction}

The holographic correspondence \cite{adscft1,adscft2,adscft3} is a powerful nonperturbative tool that has been widely used to investigate the properties of strongly coupled non-Abelian gauge theories with a large number of colors. In fact, its relevance to the physics of the strongly-coupled quark gluon plasma formed in relativistic heavy ion collisions \cite{Gyulassy:2004zy} became evident after the discovery \cite{Kovtun:2004de} that strongly coupled (spatially isotropic) plasmas that can be described by holographic methods behave as nearly perfect fluids where the shear viscosity to entropy density ratio, $\eta/s$, is close to the estimates obtained within relativistic hydrodynamic modeling of heavy ion collisions (for a recent discussion see \cite{Heinz:2013th}). Other applications of the correspondence to the physics of the Quark-Gluon Plasma (QGP) have been reviewed in \cite{CasalderreySolana:2011us}. 

Given the recent interest regarding the effects of strong electromagnetic fields in the physics of strong interactions \cite{Kharzeev:2013jha}, it is natural to investigate whether holography can also be as insightful in this case. For instance, it has been shown in \cite{Critelli:2014kra} that in a presence of a magnetic field, $\mathcal{B}$, the shear viscosity tensor of strongly coupled $\mathcal{N}=4$ SYM theory becomes anisotropic and the shear viscosity coefficient in the direction of the magnetic field violates the $\eta/s = 1/(4\pi)$ result \cite{Kovtun:2004de}.

Motivated by the recent lattice work on the effects of strong external (Abelian) magnetic fields on the QCD heavy quark potential at zero temperature done in \cite{Bonati:2014ksa}, in this paper we study the effect of a constant magnetic field on the heavy quark potential in strongly coupled $\mathcal{N}=4$ SYM theory both at zero and nonzero temperature $T$. The magnetic field distinguishes the different orientations of the $Q\bar{Q}$ pair axis with respect to direction of the magnetic field (defined here to be $z$ axis) and, thus, there is now a perpendicular potential, $V_{Q\bar{Q}}^{\perp}$, for which the pair's axis is on the transverse plane $xy$ and also a parallel potential, $V_{Q\bar{Q}}^{\parallel}$, for which the $Q\bar{Q}$ axis coincides with that of the magnetic field. Clearly, other orientations are possible but here we shall focus only on these two cases. Also, in this paper we will not solve the Schr\"odinger equation associated with this anisotropic potential (for recent studies of the effect of magnetic fields on the masses of bound states see, for instance, \cite{Machado:2013rta,Machado:2013yaa,Alford:2013jva}).

These heavy quark potentials (both at zero and nonzero temperature) in the gauge theory are defined in this paper via their corresponding identification involving the appropriate Wilson loops
\begin{equation}
\lim_{\mathcal{T}\to\infty}\langle W(\mathcal{C}_{\parallel})\rangle  \sim e^{ i V_{Q\bar{Q}}^{\parallel} \mathcal{T}   }\, \qquad \lim_{\mathcal{T}\to\infty}\langle W(\mathcal{C}_{\perp})\rangle  \sim e^{ i V_{Q\bar{Q}}^{\perp} \mathcal{T}   }\,,
\end{equation} 
where $\mathcal{C}_{\parallel}$ is a rectangular time-like contour of spatial length $L_\parallel$ in the $z$ direction and extended over $\mathcal{T}$ in the time direction while $\mathcal{C}_{\perp}$ is the corresponding contour of spatial length $L_\perp$ in the $x$ direction\footnote{Due to the matter content of $\mathcal{N} = 4$ SYM theory, the Wilson loop also contains the coupling to the six $SU(N)$ adjoint scalars $X^I$. In this paper we shall neglect the dynamics of the scalars and the holographic calculation of the Wilson loop is defined in 5 dimensions.}. We shall follow D'Hoker and Kraus' construction of the holographic dual of $\mathcal{N}=4$ SYM theory in the presence of a magnetic field \cite{DK1,DK2,DK3} and perform the calculations of the loops defined above in the background given by the asymptotic $AdS_5$ holographic Einstein-Maxwell model to be reviewed below. 

This paper is organized as follows. In the next section we review the necessary details about the holographic dual of $\mathcal{N}=4$ SYM theory in the presence of a magnetic field at zero temperature and perform the calculation of the parallel and perpendicular potentials and forces in this case. The effects of the breaking of $SO(3)$ spatial invariance induced by the magnetic field on the heavy quark potential and the the interquark force at nonzero temperature are studied in Sec.\ \ref{sec3}. Our conclusions are presented in Sec.\ \ref{conclusion} and other minor details of the calculations can be found in the Appendices \ref{apa} and \ref{apb}. We use a mostly plus metric signature and natural units $\hbar=k_B=c=1$.

\section{The holographic setup at zero temperature}
\label{sec2}

In this section we review the properties of the asymptotic $AdS_5$ background corresponding to the holographic dual of strongly coupled $\mathcal{N}=4$ SYM theory in a magnetic field worked out by D'Hoker and Kraus in \cite{DK1,DK2,DK3}. We shall focus here on the $T=0$ properties of the model.

The holographic model involves the Einstein-Maxwell action in the bulk\footnote{Our definition for the Riemann tensor has an overall minor sign in comparison to that used in \cite{DK1} and we have also changed the normalization of the Maxwell stress tensor term in the action.}:
\begin{align}
S&=\frac{1}{16\pi G_5}\int_{\mathcal{M}_5}d^5x\sqrt{-g}\left[R+\frac{12}{\ell^2}-\frac{1}{4}F_{\mu\nu}F^{\mu\nu}\right]+ S_{\textrm{GHY}}+S_{\textrm{CT}}+S_{\textrm{CS}},
\label{2.1}
\end{align}
where $\ell$ is the asymptotic AdS radius, $S_{\textrm{GHY}}$ is the Gibbons-Hawking-York action \cite{ghy1,ghy2} and $S_{\textrm{CT}}$ is the counterterm action, which is constructed using the holographic renormalization procedure \cite{ren1,ren2,ren3,ren4}. For the kind of calculations we shall carry out in the present work, these two boundary actions will not come into play and, therefore, we do not write them explicitly. The topological Chern-Simons action, $S_\textrm{CS}$, vanishes on-shell for the kind of backgrounds we shall consider here but it can be used to fix the relation between the bulk magnetic field and the magnetic field in the gauge theory by evaluating the $U(1)$ $R$-symmetry current anomaly in $4D$ $\mathcal{N}=4$ SYM\footnote{Consider writing the fields of $\mathcal{N}=4$ in terms of $\mathcal{N}=1$ with an $U(1)$ R-symmetry and assigning R-charge equals 1 to the $N^2$ gauginos \cite{DK1}.} with an external magnetic field and comparing to the variation of \eqref{2.1} under a gauge transformation, which reduces to the gauge variation of $S_{\textrm{CS}}$. Then, one concludes that the external magnetic field in SYM is $\sqrt{3}$ times the bulk magnetic field \cite{DK1}. In what follows, we discuss some of the details of the numerical evaluation of the anisotropic background at $T=0$ obtained in \cite{DK2}. Then we proceed to employ it to evaluate the parallel and perpendicular heavy quark potentials and forces in Sec.\ \ref{sec2.1} and \ref{sec2.2}, respectively.

We begin by considering the following Ansatz for the line element in light-cone coordinates\footnote{See Appendix \ref{apa}.} \cite{DK2}
\begin{align}
ds^2=\frac{dr^2}{P^2(r)}+2P(r)dudv+e^{2W(r)}(dx^2+dy^2),\,\,\,F_2=Bdx\wedge dy,
\label{2.2}
\end{align}
where the boundary of the asymptotically $AdS_5$ space is located at $r\rightarrow\infty$. A simple gauge choice for the Maxwell field giving the electromagnetic field strength tensor specified above is $A_1=Bx\,dy$. Maxwell's equations, $\nabla_\mu F^{\mu\nu}=0$, are then automatically satisfied.

The Einstein's equations obtained from the action (\ref{2.1}) are 
\begin{align}
R_{\mu\nu}-\frac{1}{2}g_{\mu\nu}R=8\pi G_5 \,T_{\mu\nu},
\label{2.3}
\end{align}
where
\begin{align}
T_{\mu\nu}\equiv\frac{-2}{\sqrt{-g}}\frac{\delta S_{\textrm{matter}}}{\delta g^{\mu\nu}} =\frac{1}{16\pi G_5}\left[g^{\alpha\beta} F_{\mu\alpha} F_{\nu\beta}-g_{\mu\nu}\left(-\frac{12}{\ell^2}+\frac{1}{4} F_{\alpha\beta}^2\right)\right],
\label{2.4}
\end{align}
is the stress-energy tensor of the gauge field $A_\mu$. After taking the trace of (\ref{2.3}) it is possible to express Einstein's equations in a more convenient form
\begin{align}
R_{\mu\nu}+\frac{g_{\mu\nu}}{3}\left[\frac{12}{\ell^2}+\frac{1}{4}F_{\alpha\beta}F^{\alpha\beta}\right] -\frac{1}{2}g^{\alpha\beta} F_{\mu\alpha} F_{\nu\beta}=0.
\label{2.7}
\end{align}
The set of linearly independent components of Einstein's equations is given by the $rr$-, $uv$- and $xx$-components of (\ref{2.7}), respectively\footnote{From here on we adopt units where $\ell=1$.}
\begin{align}
W''+\frac{P''}{2P}+W'\,^2+\frac{P'\,^2}{4P^2}+\frac{P'W'}{P}-\frac{1}{6P^2}\left(12+\frac{B^2}{2}e^{-4W}\right) &=0,\label{2.8}\\
\frac{P''}{2P}+\frac{P'\,^2}{2P^2}+\frac{P'W'}{P}-\frac{1}{3P^2}\left(12+\frac{B^2}{2}e^{-4W}\right) &=0,\label{2.9}\\
W''+2W'\,^2+\frac{2P'W'}{P}-\frac{1}{3P^2}\left(12-B^2e^{-4W}\right) &=0,\label{2.10}
\end{align}
where the prime denotes the derivative with respect to the radial direction, $r$.

Now we derive some useful equations from (\ref{2.8}), (\ref{2.9}), and (\ref{2.10}). First, we obtain a constraint by taking the combination $P^2[$(\ref{2.9})$+$(\ref{2.10})$-$(\ref{2.8})$]$
\begin{align}
P^2W'\,^2+\frac{P'\,^2}{4}+2PP'W'-\frac{1}{2}\left(12-\frac{B^2}{2}e^{-4W}\right)=0\,.
\label{2.11}
\end{align}
Taking the combinations, $2P[2$(\ref{2.8})$-$(\ref{2.9})$]$, $\frac{3P^2}{2}$(\ref{2.10})$-$(\ref{2.11}), and $2P^2e^{2W}[2$(\ref{2.9})$+$(\ref{2.10})$]$, we obtain, respectively
\begin{align}
P''+2P'W'+4P(W''+W'\,^2) &=0,\label{2.12}\\
\frac{3P^2W''}{2}+2P^2W'\,^2-\frac{P'\,^2}{4}+PP'W'+\frac{B^2}{4}e^{-4W} &=0,\label{2.13}\\
(P^2e^{2W})'' &=24e^{2W}\,.\label{2.14}
\end{align}
We shall use the coupled ODE's (\ref{2.12}) and (\ref{2.13}) to obtain the numerical solutions for $W(r)$ and $P(r)$. For this sake, we also need to specify the initial conditions to start the numerical integration of these ODE's. We are going to work with infrared boundary conditions which we shall specify in a moment. First, notice we can formally solve (\ref{2.14}) for $P^2$ as follows
\begin{align}
P^2(r)=24e^{-2W(r)}\int_0^r d\xi\int_0^\xi d\lambda e^{2W(\lambda)},
\label{2.15}
\end{align}
where we fixed the integration constants by imposing that in the infrared $P^2(0)=(P^2)'(0)=0$ \cite{DK2}. Besides (\ref{2.15}), another equation that will be useful in the determination of the parameters of the infrared expansions we shall take below for $W(r)$ and $P(r)$ is given by the combination\footnote{We note that $P(r)$ enters in this equation only through $P^2$ and $(P^2)'=2PP'$, which can be immediately read off from (\ref{2.15}).} $2[$(\ref{2.11})$+$(\ref{2.13})$]$
\begin{align}
3\left[(P^2)'W'+P^2(W''+2W'\,^2)\right]-12+B^2e^{-4W}=0\,.
\label{2.16}
\end{align}

\begin{figure}[tbp]
\begin{tabular}{cc}
\includegraphics[width=0.45\textwidth]{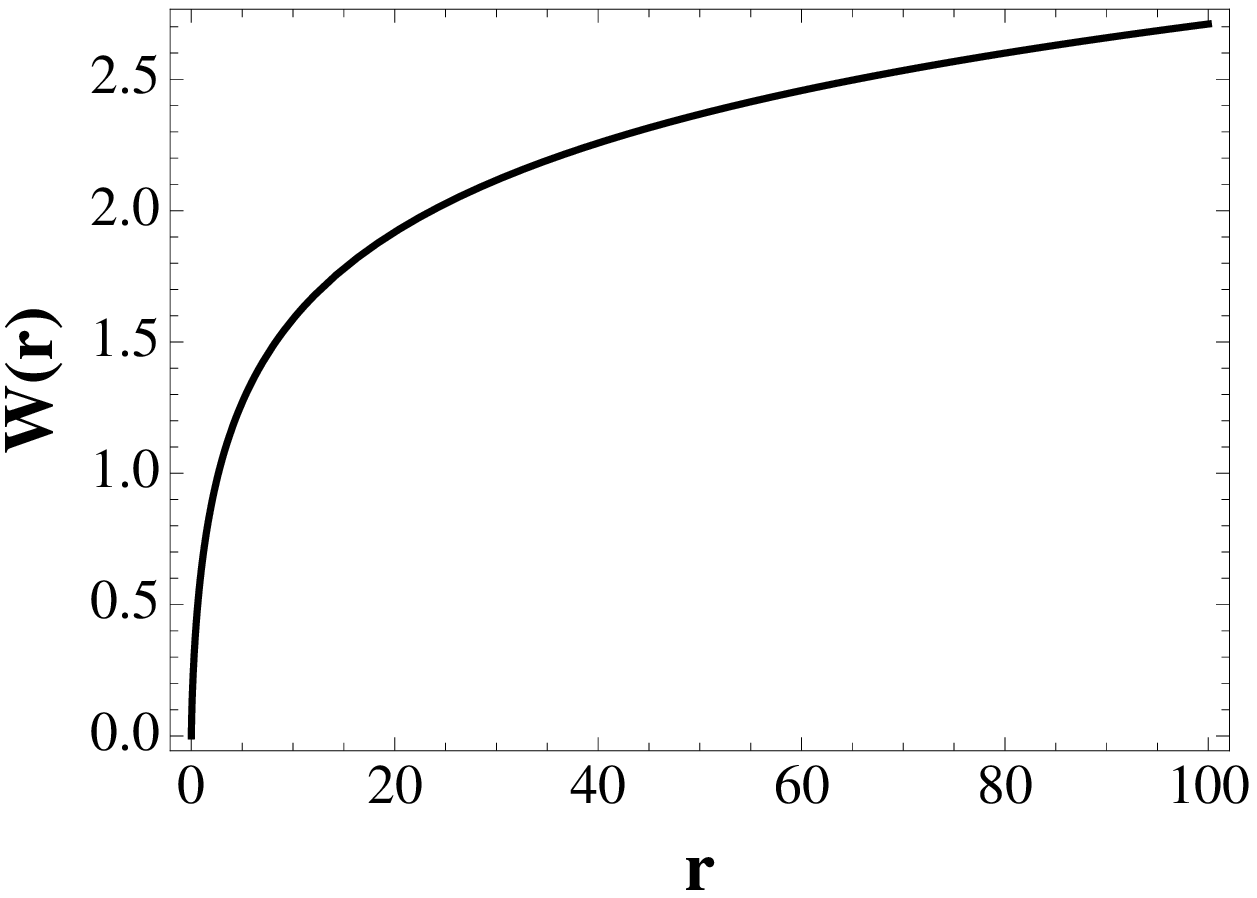} & %
\includegraphics[width=0.45\textwidth]{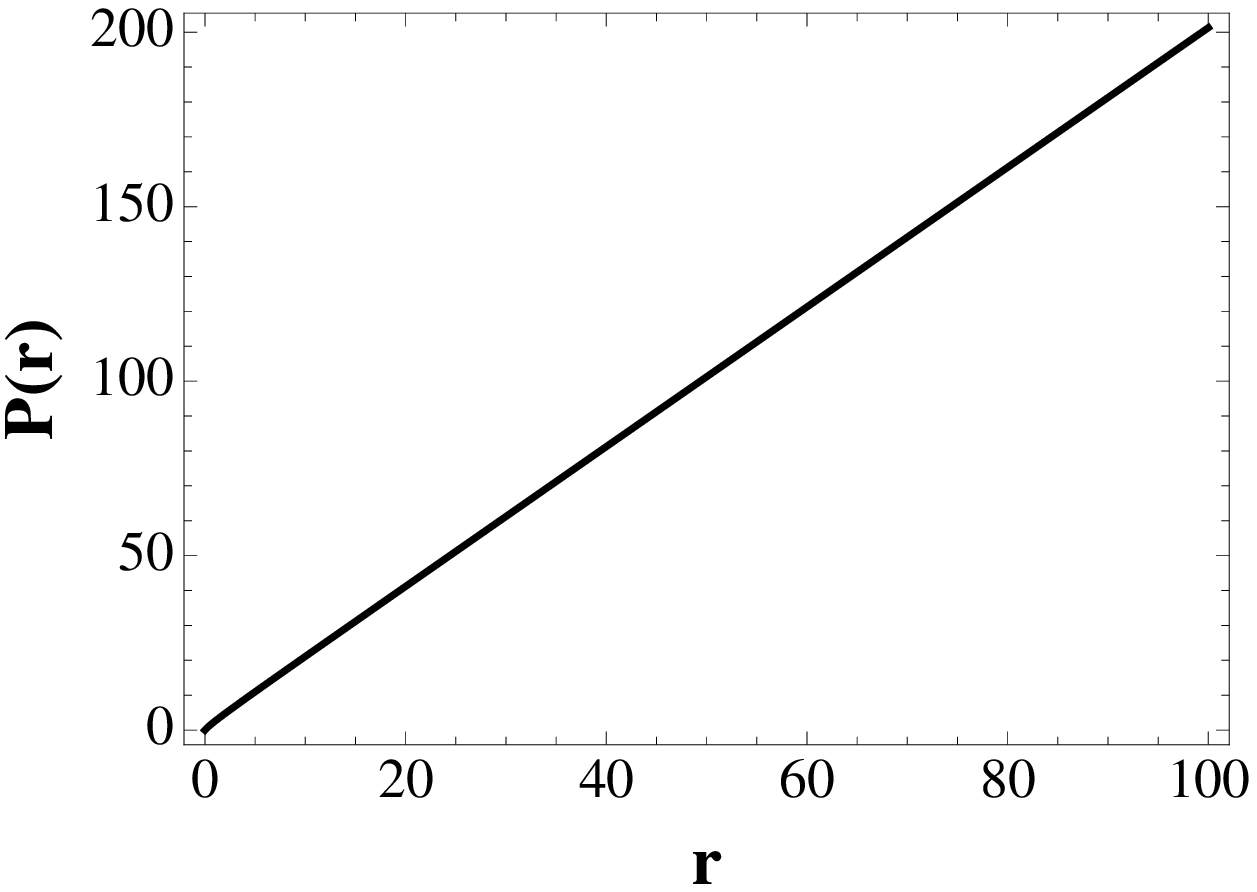} \\
&
\end{tabular}%
\caption{Numerical solution for the functions $W(r)$ and $P(r)$ that appear in the background metric at zero temperature (\ref{2.2}), which interpolates between $AdS_3\times\mathbb{R}^2$ in the infrared (small $r$) and $AdS_5$ in the ultraviolet (large $r$).}
\label{fig1}
\end{figure}

Let us now work out the infrared expansions for $W(r)$ and $P(r)$. Following \cite{DK2}, we are interested in numerical solutions of the dynamical ODE's (\ref{2.12}) and (\ref{2.13}) that interpolate between $AdS_3\times\mathbb{R}^2$ for small $r$ in the infrared and $AdS_5$ for large $r$ in the ultraviolet. As discussed in \cite{DK1}, this corresponds to a renormalization group flow between a CFT in $(1+1)$-dimensions in the infrared and a CFT in $(3+1)$-dimensions in the ultraviolet, which is the expected behavior of SYM theory in the presence of a constant magnetic field \cite{DK1}. Then, for small $r$ we can take the following infrared expansions
\begin{align}
W(r) &= r^a+\omega r^{2a}+\mathcal{O}(r^{3a}),\label{2.17}\\
P^2(r) &\approx 12r^2\left[1-2r^a+(2-2\omega)r^{2a}\right]\left[1+\frac{4r^a}{2+3a+a^2} +\frac{2(1+\omega)r^{2a}}{1+3a+2a^2}\right],\label{2.18}
\end{align}
where (\ref{2.18}) was obtained by substituting (\ref{2.17}) into (\ref{2.15}). Now we substitute (\ref{2.17}) and (\ref{2.18}) into (\ref{2.16}) and set to zero the coefficients of each power of $r$ in the resulting expression, obtaining
\begin{align}
\mathcal{O}(r^0)&:\,\,\,B=2\sqrt{3},\label{2.19}\\
\mathcal{O}(r^a)&:\,\,\,9a^2+9a-B^2=0\Rightarrow a=a_+\approx 0.758,\label{2.20}\\
\mathcal{O}(r^{2a})&:\,\,\,\omega\approx -0.634,\label{2.21}
\end{align}
where we have chosen the positive root in (\ref{2.20}) in order to obtain a finite $W(0)$ and used (\ref{2.19}) and (\ref{2.20}) to obtain (\ref{2.21}). Substituting (\ref{2.20}) and (\ref{2.21}) into (\ref{2.17}) and (\ref{2.18}), we determine the first terms in the infrared expansions for $W(r)$, $W'(r)$, $P(r)$ and $P'(r)$, which are enough to initialize the numerical integration of the coupled ODE's (\ref{2.12}) and (\ref{2.13}). We start the integration in the deep infrared at some small $r=r_{\textrm{min}}$ and integrate up to some large $r=r_{\textrm{max}}$ near the boundary. The numerical results for the metric functions $W(r)$ and $P(r)$ appearing in (\ref{2.2}) are shown in Fig.\ \ref{fig1} (these results match those in \cite{DK2}).

The ultraviolet asymptotics for this numerical solution is given by: $\left(e^{2W(r_\textrm{max})},P(r_\textrm{max})\right)$ $\approx\left(1.12365,1.00002\right)\times 2r_\textrm{max}$. Therefore, in order to have an asymptotically $AdS_5$ space at the ultraviolet cutoff, $r=r_\textrm{max}$, we rescale $\left(e^{2W(r)},P(r)\right)\mapsto \left(e^{2\bar{W}(r)},\bar{P}(r)\right)$, where $e^{2\bar{W}(r)}=e^{2W(r)}/1.12365$ and $\bar{P}(r)=P(r)/1.00002$. With this metric rescaling, the physical constant magnetic field in the gauge theory reads\footnote{This rescaling changes the $x$- and $y$-coordinates in (\ref{2.2}) as follows: $\left(x,y\right)\mapsto\left(x,y\right)/\sqrt{1.12365}$. Furthermore, as discussed after eq. (\ref{2.1}), the extra factor of $\sqrt{3}$ relates the bulk magnetic field and the magnetic field in the gauge theory.}: $\mathcal{B}=\sqrt{3}B/1.12365\approx 5.34$.

These results were originally obtained in Ref.\ \cite{DK2}. In the following we use them to evaluate the parallel and perpendicular heavy quark potential at zero temperature in the presence of a constant magnetic field.

\subsection{Holographic Wilson loop $\parallel \v{\mathcal{B}}$ at $T=0$}
\label{sec2.1}

Now we determine the parallel heavy quark potential from the VEV of a rectangular Wilson loop defined by a contour $\mathcal{C}_\parallel$ with its spatial length along the magnetic field direction. We follow the holographic prescription proposed in \cite{maldacena,Rey:1998ik,rey,sonne} (see also \cite{yaffe,jorge} and references therein for more recent discussions) to evaluate the rectangular loops in SYM in the strong t'Hooft coupling limit, $\lambda \gg 1$, with a large number of colors, $N \to \infty$, in terms of a classical Nambu-Goto action in the background discussed in the previous section. 

For this sake, it is better to recast the rescaled version of the metric (\ref{2.2}) as follows\footnote{See Appendix \ref{apa}.}
\begin{align}
ds^2=\frac{dr^2}{\bar{P}^2(r)}+\bar{P}(r)(-dt^2+dz^2)+e^{2\bar{W}(r)}(dx^2+dy^2),
\label{2.22}
\end{align}
where $\bar{P}(r)$ and $\bar{W}(r)$ are the rescaled numerical functions discussed in the previous section. For the sake of notation simplicity, since in the remaining of this section we are going to use only these rescaled functions, we shall omit  from now on the bars in their notation.

The rectangular Wilson loop at the boundary of the asymptotically $AdS_5$ space (\ref{2.22}) parallel to the magnetic field  is extended along the time direction by $\mathcal{T}$ and has spatial length $L^{\parallel}$, which denotes the heavy quark-antiquark spatial separation in the direction of the magnetic field (we take $\mathcal{T}\gg L^{\parallel}$). We choose to place the probe quark $Q$ at $-\hat{z}L^{\parallel}/2$ and the $\bar{Q}$-probe charge at $+\hat{z}L^{\parallel}/2$. Attached to each of the probe charges in the pair there is a string that sags in the interior of the bulk of the space (\ref{2.22}). As usual \cite{maldacena,Rey:1998ik,rey,sonne}, in the limit $\mathcal{T}\to \infty$ we consider a classical U-shaped configuration that extremizes the Nambu-Goto action and has a minimum at some value $r_0$ of the radial coordinate in the interior of the bulk. 

The parametric equation of the 2-dimensional string worldsheet swept out in the 5-dimensional bulk is formally given by
\begin{align}
X^\mu &: \textrm{Internal Space} \rightarrow \textrm{Target Space (Bulk)}\nonumber\\
      & \,\,\,\,\,\,\,\,\,\,\,\,\,\,\,\,\,\,\,\,\,\,\,\,\,\,\,\,\, (\tau,\sigma) \mapsto X^\mu(\tau,\sigma) = x^\mu,
\label{2.23}
\end{align}
and, in static gauge $\tau\rightarrow t,\,\,\, \sigma\rightarrow z$, the target space coordinates over the string worldsheet become
\begin{align}
X^r(t,z)=r,\,\,\,X^t=t,\,\,\,X^x=0,\,\,\,X^y=0,\,\,\,X^z=z,
\label{2.25}
\end{align}
where $X^r(t,z)=r$ is a constraint equation. For loops where $\mathcal{T} \gg L^{\parallel}$, the static string configuration is invariant under translations in time and one can write $X^r(t,z)=X^r(z)=r$. For the sake of notation simplicity, we take a slight abuse of language and write simply $r=r(z)$ for this constraint equation. Therefore, the static gauge condition can be summarized as follows
\begin{align}
(\tau,\sigma)\rightarrow (t,z)\Rightarrow X^\mu(t,z)=(r(z),t,0,0,z).
\label{2.26}
\end{align}

The pullback or the induced metric over the string worldsheet in the numerical background (\ref{2.22}) is defined by
\begin{align}
\gamma_{ab}=g_{\mu\nu}\partial_a X^\mu\partial_b X^\nu,\,\,\,a,b\in\{\tau,\sigma\},
\label{2.27}
\end{align}
with components 
\begin{align}
\gamma_{tz}&=\gamma_{zt}=0,\label{2.28}\\
\gamma_{tt}&=-P(r(z)),\label{2.29}\\
\gamma_{zz}&=\frac{\dot{r}^2(z)}{P^2(r(z))}+P(r(z)),\label{2.30}
\end{align}
where the dot denotes the derivative with respect to $z$. The square root of minus the determinant of the induced metric reads
\begin{align}
\sqrt{-\gamma}=\sqrt{\frac{\dot{r}^2(z)}{P(r(z))}+P^2(r(z))},
\label{2.31}
\end{align}
and, therefore, the Nambu-Goto action for this $Q\bar{Q}$-configuration is
\begin{align}
S_{\textrm{NG}}=\frac{1}{2\pi\alpha'}\int d^2\sigma \sqrt{-\gamma}= \frac{\mathcal{T}}{2\pi \alpha'} \int_{-L^{\parallel}/2}^{L^{\parallel}/2} dz \sqrt{\frac{\dot{r}^2(z)}{P(r(z))}+P^2(r(z))}\,,
\label{2.32}
\end{align}
where $\alpha'=\ell_s^2$ and $\ell_s$ is the string length.

Since the integrand in (\ref{2.32}), $L_{\textrm{NG}}$, does not depend explicitly on $z$, $H_{\textrm{NG}}$ defined below is a \emph{constant of motion} in the z direction
\begin{align}
H_{\textrm{NG}}\equiv \frac{\partial L_{\textrm{NG}}}{\partial\dot{r}}\dot{r}-L_{\textrm{NG}}=\frac{-P^2(r(z))}{\sqrt{\frac{\dot{r}^2(z)}{P(r(z))}+P^2(r(z))}}=C\,.
\label{2.33}
\end{align}
We may determine $C$ by evaluating (\ref{2.33}) at the minimum of $r(z)$ where the U-shaped string configuration has a minimum in the interior of the bulk, $r(z=0)=r_0$, where $\dot{r}(0)=0$ and find
\begin{align}
C = \frac{-P^2(r_0)}{\sqrt{P^2(r_0)}}\,.
\label{2.34}
\end{align}
Substituting (\ref{2.34}) into the square of (\ref{2.33}) and solving for $\dot{r}(z)$, one obtains 
\begin{align}
\dot{r}(z)=\frac{dr(z)}{dz}=\sqrt{P^3(r(z))\left[\frac{P^2(r(z))}{P^2(r_0)}-1\right]},
\label{2.35}
\end{align}
which implies that
\begin{align}
L^\parallel(r_0)= 2\int_{r_0}^{\infty} \frac{dr}{\sqrt{P^3(r)\left[\frac{P^2(r)}{P^2(r_0)}-1\right]}},
\label{2.36}
\end{align}
where we used that for the U-shaped string configuration described before, $r(\pm L^{\parallel}/2)\rightarrow\infty$, since the probe charges are localized at the boundary of the space (\ref{2.22}), and we also took into account the fact that the U-shaped contour of integration in the $rz$-plane is symmetric with respect to the $r$-axis, with $r(z=0)=r_0$.

The bare parallel heavy quark potential for this static $Q\bar{Q}$-configuration reads
\begin{align}
V_{Q\bar{Q},\textrm{bare}}^{\parallel}(r_0)=\frac{S_{\textrm{NG}}}{\mathcal{T}}\biggr|_{\textrm{on-shell}} &=\frac{1}{2\pi\alpha'}\int_{-L^{\parallel}/2}^{L^{\parallel}/2} dz \sqrt{\frac{P^4(r(z))}{P^2(r(0))}}\nonumber\\
&=\frac{1}{\pi \alpha'}\int_{r_0}^\infty dr\sqrt{\frac{P(r)}{P^2(r)-P^2(r_0)}},
\label{2.37}
\end{align}
where we used (\ref{2.35}) to evaluate the on-shell Nambu-Goto action (\ref{2.32}). Now we need to regularize (\ref{2.37}) by subtracting the divergent self-energies of the infinitely heavy probe charges $Q$ and $\bar{Q}$. These contributions correspond to strings stretching from each probe charge at the boundary to the deep interior of the bulk and, in practice, one identifies the ultraviolet divergences to be subtracted by looking at the dominant contribution in the integrand of (\ref{2.37}) in the limit $r\rightarrow\infty$
\begin{align}
\sqrt{\frac{P(r)}{P^2(r)-P^2(r_0)}}\stackrel{r\rightarrow\infty}{\longrightarrow} \frac{1}{\sqrt{P(r)}}\biggr|_{r\rightarrow\infty}\sim\frac{1}{\sqrt{2r}}\,.
\label{2.38}
\end{align}
Therefore, the sum of the self-energies of the probe charges is given by
\begin{align}
2\times V_0=2\times\frac{1}{2\pi \alpha'}\int_0^\infty\frac{dr}{\sqrt{2r}},
\label{2.39}
\end{align}
and the renormalized parallel heavy quark potential is
\begin{align}
V_{Q\bar{Q}}^{\parallel}(r_0)=V_{Q\bar{Q},\textrm{bare}}^{\parallel}(r_0)-2V_0= \frac{1}{\pi \alpha'}\left[ \int_{r_0}^\infty dr\left(\sqrt{\frac{P(r)}{P^2(r)-P^2(r_0)}}-\frac{1}{\sqrt{2r}}\right) - \int_0^{r_0}\frac{dr}{\sqrt{2r}}\right]\,.
\label{2.40}
\end{align}

In order to obtain the curve $V_{Q\bar{Q}}^{\parallel}(L^{\parallel})$, one may construct a table with pairs of points $(L^{\parallel}(r_0),V_{Q\bar{Q}}^{\parallel}(r_0))$ by taking different values of the parameter $r_0$ in Eqs.\ (\ref{2.36}) and (\ref{2.40}), and then numerically interpolate between these points. Before doing this, let us first obtain the corresponding expressions for the perpendicular potential $V_{Q\bar{Q}}^\perp(L^\perp)$. After that, we will make a comparison between the heavy quark potentials and forces obtained in the presence of the magnetic field and the standard isotropic SYM results discussed in \cite{maldacena}.

\subsection{Holographic Wilson loop $\perp \v{\mathcal{B}}$ at $T=0$}
\label{sec2.2}

Now we consider a rectangular Wilson loop with spatial length $L^\perp$ located in the plane perpendicular to the magnetic field direction at the boundary of the space (\ref{2.22}). We place the $Q$-probe charge at $-\hat{x}L^{\perp}/2$ and the $\bar{Q}$-probe charge at $+\hat{x}L^{\perp}/2$. For this $Q\bar{Q}$-configuration, it is convenient to define the following static gauge
\begin{align}
(\tau,\sigma)\rightarrow (t,x)\Rightarrow X^\mu(t,x)=(r(x),t,x,0,0)\,.
\label{2.41}
\end{align}

Following the same general steps discussed in detail in the previous section, one obtains
\begin{align}
L^{\perp}(r_0)&= 2\int_{r_0}^{\infty} \frac{dr}{\sqrt{P^2(r)e^{2W(r)}\left[\frac{P(r)e^{2W(r)}}{P(r_0)e^{2W(r_0)}}-1\right]}},\label{2.42}\\
V_{Q\bar{Q}}^{\perp}(r_0)&= \frac{1}{\pi \alpha'}\left[ \int_{r_0}^\infty dr\left(\sqrt{\frac{e^{2W(r)}}{P(r)e^{2W(r)}-P(r_0)e^{2W(r_0)}}}-\frac{1}{\sqrt{2r}}\right) - \int_0^{r_0}\frac{dr}{\sqrt{2r}}\right].\label{2.43}
\end{align}
We note that both the (renormalized) parallel and perpendicular potentials are regularized by the same subtraction term, $2\times V_0$, in Eq.\ (\ref{2.39}).

\begin{figure}[tbp]
\begin{center}
\begin{tabular}{cc}
\includegraphics[width=0.45\textwidth]{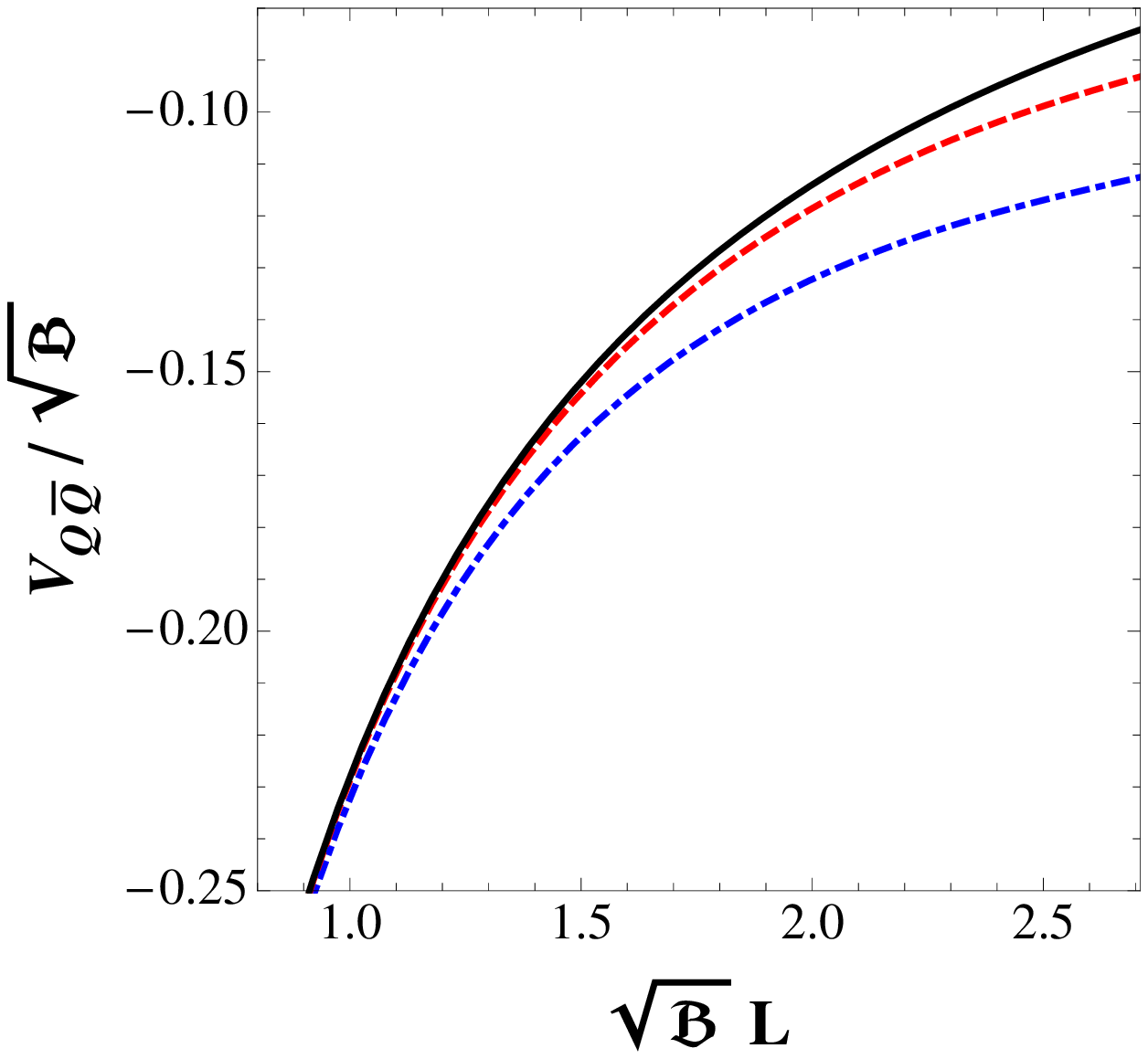} & %
\includegraphics[width=0.45\textwidth]{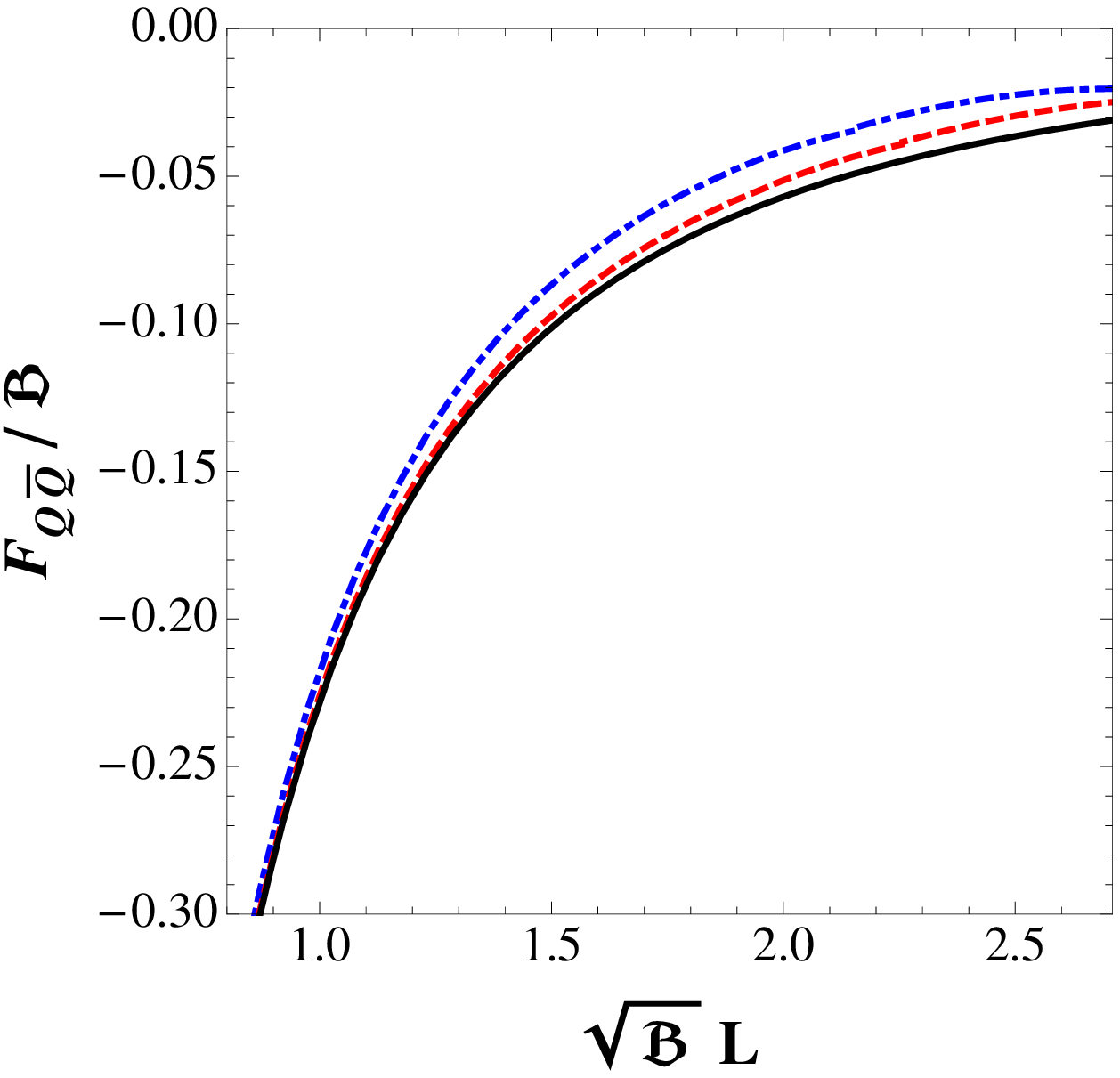} \\
&
\end{tabular}%
\caption{(Color online) Anisotropy induced by a magnetic field $\mathcal{B}$ in $\mathcal{N}=4$ SYM at $T=0$ (in this plot $\alpha'=1$) in the heavy quark potential (left panel) and the corresponding force (right panel). The solid black lines denote the isotropic result $\sim -0.228/L$ \cite{maldacena}, the dashed red lines correspond to the perpendicular potential $V_{Q\bar{Q}}^\perp$ and force $F_{Q\bar{Q}}^\perp=-dV_{Q\bar{Q}}^\perp/dL$, and the dotted-dashed blue lines correspond to the parallel potential $V_{Q\bar{Q}}^\parallel$ and force $F_{Q\bar{Q}}^\parallel=-dV_{Q\bar{Q}}^\parallel/dL$.}
\label{fig2}
\end{center}
\end{figure}

In practice, for the numerical integrations to be performed in Eqs.\ (\ref{2.36}), (\ref{2.40}), (\ref{2.42}), and (\ref{2.43}), the boundary at $r\rightarrow\infty$ is numerically described by $r_{\textrm{max}}$, in accordance with the numerical solution obtained for the metric (\ref{2.22}). Our plots for the parallel and perpendicular potentials at $T=0$ are shown on the left panel of Fig.\ \ref{fig2}. One can see that for the $T=0$ anisotropic holographic setup considered in this section the magnitudes of both the parallel and perpendicular potentials at nonzero $\mathcal{B}$ are enhanced with respect to the $\mathcal{B}=0$ isotropic case (given by $\sim -0.228/L$ \cite{maldacena}), though the parallel potential is more affected by the magnetic field. Also, for very short distances $\sqrt{\mathcal{B}}\, L \ll 1$, both potentials converge to the isotropic potential \cite{maldacena} since the effects from the magnetic field become negligible in this limit. On the right panel of Fig.\ \ref{fig2} we show the forces associated with these potentials. One can see that the magnetic field generally decreases the magnitude of the attractive force between the quarks in comparison to the isotropic scenario and that the force experienced by the quarks becomes the weakest when the pair axis is parallel to the direction of the magnetic field. 

Moreover, in the absence of any other scale in the theory besides $\mathcal{B}$ (and the interquark distance $L$), the actual value of $\mathcal{B}$ is immaterial. This situation changes when one switches on the temperature and, in this case, there is a new dimensionless scale given by the ratio $\mathcal{B}/T^2$. In fact, we shall see in the next section that in this case one is able to tune the anisotropy in the heavy quark potential by varying the value of the magnetic field.

\section{The holographic setup at finite temperature}
\label{sec3}

In this section we study the interplay between finite temperature and magnetic field effects on the heavy quark potential in an $\mathcal{N}=4$ SYM plasma. For the sake of completeness, here we review some of the details regarding the derivation of the numerical anisotropic metric at finite temperature obtained in \cite{DK1}. We then proceed to employ it to determine how the parallel and perpendicular potentials are affected by the magnetic field and temperature.

We use the following Ansatz\footnote{See Appendix \ref{apa} for a discussion about the coordinates used in this section.} \cite{DK1}
\begin{align}
ds^2=\frac{d\bar{r}^2}{h(\bar{r})}-h(\bar{r})dt^2+e^{2W(\bar{r})}(dx^2+dy^2)+e^{2G(\bar{r})}dz^2,\,\,\,F_2=Bdx\wedge dy\,,
\label{3.1}
\end{align}
where the boundary of the asymptotically $AdS_5$ space is taken to be at $\bar{r}\rightarrow\infty$ and the black hole horizon is located at $\bar{r}=\bar{r}_H$, which is defined by the largest root of the equation $h(\bar{r}_H)=0$. Moreover, note that $W(\bar{r}_H)$ and $G(\bar{r}_H)$ are both finite.

The set of linearly independent components of Einstein's equations are given by the $\bar{r}\bar{r}$-, $tt$-, $xx$- and $zz$-components of (\ref{2.7}), respectively
\begin{align}
G''+G'\,^2+2W''+2W'\,^2+\frac{h''}{2h}+\frac{h'}{h}\left(W'+\frac{G'}{2}\right)-\frac{4}{h} \left(1+\frac{B^2}{24}e^{-4W}\right) &=0,\label{3.2}\\
h''+h'(G'+2W')-8\left(1+\frac{B^2}{24}e^{-4W}\right) &=0,\label{3.3}\\
W''+2W'\,^2+W'\left(\frac{h'}{h}+G'\right)-\frac{4}{h}\left(1-\frac{B^2}{12}e^{-4W}\right) &=0,\label{3.4}\\
G''+G'\,^2+G'\left(\frac{h'}{h}+2W'\right)-\frac{4}{h}\left(1+\frac{B^2}{24}e^{-4W}\right) &=0\,.
\label{3.5}
\end{align}

Now we derive some useful equations from (\ref{3.2}), (\ref{3.3}), (\ref{3.4}), and (\ref{3.5}). First, we obtain a constraint by taking the combination $-h[$(\ref{3.2})$-\frac{1}{2h}$(\ref{3.3})$-2$(\ref{3.4})$-$(\ref{3.5})$]$
\begin{align}
h'(G'+2W')+2h(W'\,^2+2W'G')-12\left(1-\frac{B^2}{24}e^{-4W}\right)=0\,.
\label{3.6}
\end{align}
Taking the combination $h[$(\ref{3.4})$-$(\ref{3.5})$]$, and using the constraint (\ref{3.6}) to eliminate $B^2e^{-4W}$ from Eqs.\ (\ref{3.3}), (\ref{3.4}) and (\ref{3.5}), we obtain, respectively
\begin{align}
h(W''-G''+2W'\,^2-G'\,^2-G'W')+h'(W'-G')+\frac{B^2}{2}e^{-4W} &=0,\label{3.7}\\
h''+\frac{5(G'+2W')}{3}h'+\frac{4(W'\,^2+2W'G')}{3}h-16 &=0,\label{3.8}\\
W''+\frac{2}{3}W'\,^2-\frac{1}{3}\left(\frac{h'}{h}+5G'\right)W'+\frac{12-2h'G'}{3h} &=0,\label{3.9}\\
G''+G'\,^2+\frac{2}{3}\left(\frac{2h'}{h}+5W'\right)G'+\frac{2W'\,^2}{3}+\frac{2h'W'-24}{3h} &=0\,.\label{3.10}
\end{align}
Eqs.\ (\ref{3.6}) and (\ref{3.7}) will be employed in the determination of the near-horizon boundary conditions required to initialize the numerical integration of the coupled ODE's (\ref{3.8}), (\ref{3.9}), and (\ref{3.10}).

Following \cite{DK1}, it is convenient to rescale the radial coordinate in such a way that the horizon is at $1$, i.e., $\bar{r}\mapsto\tilde{r}$, such that $\tilde{r}_H=1$. Therefore, after this rescaling $h(1)=0$. One can also rescale the time coordinate \cite{DK1} so that
\begin{align}
h'(1)=1\,,
\label{3.12}
\end{align}
and the Hawking temperature, therefore, reads
\begin{align}
T=\frac{\sqrt{-g'_{\tilde{t}\tilde{t}}\,g^{\tilde{r}\tilde{r}}\,'}}{4\pi}\biggr|_{\tilde{r}=1}= \frac{h'(1)}{4\pi}=\frac{1}{4\pi}\,.
\label{3.13}
\end{align}
Rescaling $x$, $y$ and $z$, one can also set $W(1)=G(1)=0$ and it is possible to show that
\begin{align}
W'(1)=4-\frac{b^2}{3},\,\,\,G'(1)=4+\frac{b^2}{6}\,,
\label{3.15}
\end{align}
where $b$ is the magnetic field expressed in these rescaled coordinates. The numerical solutions will asymptote to
\begin{align}
h(\tilde{r}\rightarrow\infty,b)\rightarrow \tilde{r}^2,\,\,\,e^{2W(\tilde{r}\rightarrow\infty,b)}\rightarrow \omega(b)\tilde{r}^2, \,\,\,e^{2G(\tilde{r}\rightarrow\infty,b)}\rightarrow g(b)\tilde{r}^2,
\label{3.16}
\end{align}
where $\omega(b)$ and $g(b)$ are functions that can be determined numerically. The conformal boundary metric, therefore, reads
\begin{align}
ds^2_{\textrm{bdy}}\sim -d\tilde{t}^{\,2}+\omega(b)(d\tilde{x}^2+d\tilde{y}^2)+g(b)d\tilde{z}^2,\,\,\, F_2=b\,d\tilde{x}\wedge d\tilde{y}\,,
\label{3.17}
\end{align}
where $\sim$ denotes conformal equivalence. One can write the boundary metric (\ref{3.17}) in the usual (conformal) Minkowski form by rescaling the spatial coordinates in order to absorb the factors of $\omega(b)$ and $g(b)$ such that the physical magnetic field in the gauge theory is \cite{DK1}
\begin{align}
\mathcal{B}=\sqrt{3}\frac{b}{\omega(b)}\,.
\label{3.18}
\end{align}

Now we work out the near-horizon expansions for\footnote{For the sake of notation simplicity, we omit from now on the tilde in the rescaled coordinates.} $h(r)$, $W(r)$, and $G(r)$. In order to avoid the singular point of the ODE's located at the horizon, one must start the numerical integration slightly above it. For this sake, we take near-horizon Taylor expansions $X(r) = X(1) + X'(1) (r-1) + \cdots$, with some small but nonzero $(r-1) \rightarrow r_{\textrm{min}}$, where $X = \{h,W,G\}$. Since now $X(1) = 0$, the near-horizon boundary conditions are given by $X(r_{\textrm{start}} \equiv r_{\textrm{min}} + 1) \approx X'(1) r_{\textrm{min}}$, with the numerical integration starting at $r_{\textrm{start}}=r_{\textrm{min}} + 1$, slightly above the horizon, and going up to some $r_{\textrm{max}}$ near the boundary. The near-horizon boundary conditions are then
\begin{align}
&h(r_{\textrm{start}}) = r_{\textrm{min}},\,\,\,h'(r_{\textrm{start}})=1,\nonumber\\
&W(r_{\textrm{start}}) = \left(4-\frac{b^2}{3}\right) r_{\textrm{min}},\,\,\, W'(r_{\textrm{start}}) = 4-\frac{b^2}{3}\,,\nonumber\\
&G(r_{\textrm{start}}) = \left(4+\frac{b^2}{6}\right) r_{\textrm{min}},\,\,\, G'(r_{\textrm{start}}) = 4+\frac{b^2}{6}\,.
\label{3.19}
\end{align}

\begin{figure}[tbp]
\begin{center}
\includegraphics[width=0.45\textwidth]{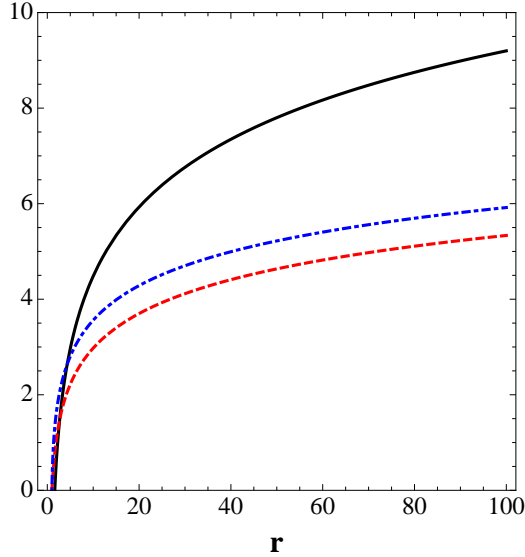}
\caption{(Color online) Numerical solutions for the functions $\ln\,h(r)$ (solid black line), $W(r)$ (dashed red line), and $G(r)$ (dotted-dashed blue line) in (\ref{3.1}) for $b=2.7$. The numerical background interpolates between $\textrm{BTZ}\times\mathbb{R}^2$ in the infrared (small $r$) and $AdS_5$ in the ultraviolet (large $r$).}
\label{fig3}
\end{center}
\end{figure}

The numerical results for $\ln\, h(r)$, $W(r)$, and $G(r)$ in the metric (\ref{3.1}) are shown in Fig.\ \ref{fig3} for $b=2.7$. The numerical solutions interpolate between $\textrm{BTZ}\times\mathbb{R}^2$ in the infrared and $AdS_5$ in the ultraviolet\footnote{The BTZ solution \cite{btz} is asymptotically $AdS_3$ and, thus, at zero temperature one recovers the solution discussed in Sec.\ \ref{sec2}.}. These results were originally obtained in \cite{DK1} and we have checked that we can numerically reproduce the entropy density found in that work.

We remark that for $b\ge 2\sqrt{3}$ the numerical solutions for the geometry are not asymptotically $AdS_5$ \cite{DK1}. Thus, we restrict our calculations to values of $b<2\sqrt{3}$. It is important to emphasize, however, that this limitation on the values of $b$ does not imply in any practical limitation on the values of the physical magnetic field in the gauge theory, $\mathcal{B}$, which is related to $b$ via eq. (\ref{3.18}). This is so because, as one can check numerically, $\omega(b)$ is a decreasing function of $b$ and $\omega(b\rightarrow 2\sqrt{3})\rightarrow 0$, such that one can cover in practice all values of the physical magnetic field in the interval $[0,\infty)$.

Also, as in the zero temperature case, in order to have an asymptotically $AdS_5$ space at the ultraviolet cutoff, $r=r_{\textrm{max}}$, we see from (\ref{3.16}) that one needs to do the following rescaling: $\left(e^{2W(r,b)},e^{2G(r,b)}\right)\mapsto\left(e^{2\bar{W}(r,b)},e^{2\bar{G}(r,b)}\right)$, where $e^{2\bar{W}(r,b)}=e^{2W(r,b)}/\omega(b)$ and $e^{2\bar{G}(r,b)}=e^{2G(r,b)}/g(b)$. For the sake of notation simplicity, since in the remaining of this work we are going to use only the rescaled functions $e^{2\bar{W}(r,b)}$ and $e^{2\bar{G}(r,b)}$, we shall omit from now on the bars in their notation. We can now use this background to evaluate the parallel and perpendicular heavy quark potentials in an $\mathcal{N}=4$ SYM plasma in the presence of a constant magnetic field.

\subsection{Anisotropic heavy quark potential for $T\neq 0$}
\label{sec3.1}

The holographic calculation of the $T\neq 0$ Wilson loops used in the definition of the parallel and perpendicular potentials follows the same procedure done before in the case where $T=0$. The boundary conditions for each string configuration are the same as before and the overall shape of the string in the bulk is the U-shaped profile \cite{sonne}. The only difference is that when $T\neq 0$ the background metric to be used is the numerically found anisotropic black brane in Eq.\ (\ref{3.1}) according to the discussion above. Therefore, it is easy to show that 
the interquark separation and (renormalized) heavy quark potential for the parallel case are 
\begin{align}
L^{\parallel}(r_0)&= 2\int_{r_0}^{\infty} \frac{dr}{\sqrt{h(r)e^{2G(r)}\left[\frac{h(r)e^{2G(r)}}{h(r_0)e^{2G(r_0)}}-1\right]}},\label{3.20}\\
V_{Q\bar{Q}}^{\parallel}(r_0)&= \frac{1}{\pi\alpha'}\left[ \int_{r_0}^\infty dr\left(\sqrt{\frac{h(r)e^{2G(r)}}{h(r)e^{2G(r)}-h(r_0)e^{2G(r_0)}}}-1\right) - \int_0^{r_0}dr\right]\,
\label{3.21}
\end{align}
while for the perpendicular setup one finds
\begin{align}
L^{\perp}(r_0)&= 2\int_{r_0}^{\infty} \frac{dr}{\sqrt{h(r)e^{2W(r)}\left[\frac{h(r)e^{2W(r)}}{h(r_0)e^{2W(r_0)}}-1\right]}},\label{3.22}\\
V_{Q\bar{Q}}^{\perp}(r_0)&= \frac{1}{\pi\alpha'}\left[ \int_{r_0}^\infty dr\left(\sqrt{\frac{h(r)e^{2W(r)}}{h(r)e^{2W(r)}-h(r_0)e^{2W(r_0)}}}-1\right) - \int_0^{r_0}dr\right]\,,
\label{3.23}
\end{align}
where $r_0$ is the point in the bulk where the U-shaped configuration has its minimum. Note that we used the same (temperature independent) subtraction scheme employed at $T=0$ to define the renormalized potentials at finite temperature. These potentials are proportional to the (regularized) area of the Nambu-Goto worldsheet and they are interpreted in the strongly coupled gauge theory as the difference in the total free energy of the system due to the addition of the heavy $Q\bar{Q}$-pair \cite{Noronha:2010hb}.  While one can may argue that one should remove an ``entropy-like" contribution from this free energy difference \cite{mocsy,Noronha:2009ia}, in this paper we shall not perform such a subtraction and, for simplicity, we define this free energy difference (which equals the regularized Nambu-Goto action) in each case to be the corresponding heavy quark potential at finite temperature. 

As done before, in the numerical integrations to be performed in Eqs.\ (\ref{3.20}), (\ref{3.21}), (\ref{3.22}), and (\ref{3.23}), the boundary at $r\rightarrow\infty$ is numerically described by $r_{\textrm{max}}$. At finite temperature, there is a maximum value of $LT$ above which there are other string configurations that may contribute to the evaluation of the Wilson loops at finite temperature \cite{yaffe} besides the semi-classical U-shaped string configuration. This implies that one cannot compute the potentials with the setup described here when $LT$ is large. In fact, one can show that the inclusion of the magnetic field makes this problem worse, as it is shown in Fig.\ \ref{fig4} below. In this plot we show $LT$ as a function of  the appropriate rescaled horizon $y_H$ (see Appendix \ref{apb} for the definition of this variable) for the isotropic case (solid black line) and for the parallel (dotted-dashed blue line) and perpendicular (dashed red curve) cases computed using $\mathcal{B}/T^2=50$ (left panel) and $\mathcal{B}/T^2=1000$ (right panel). When $y_h \to 0$ the curves follow the isotropic SYM case while one can see that the maximum of $LT$ is considerably decreased if the magnetic field is sufficiently intense and this effect is stronger for the perpendicular configuration. This implies that the region of applicability of the U-shaped string worldsheet decreases with the applied magnetic field and, thus, other string configurations must be taken into account when computing the string generating functional for sufficiently large $LT$ \cite{yaffe}. This problem was investigated in the case of an isotropic $\mathcal{N}=4$ SYM plasma in \cite{Grigoryan:2011cn} but the extension of these calculations to the anisotropic scenario studied here will be left as a subject of a future study. Nevertheless, for the values of $LT$ in which the U-shaped configuration is dominant our results for the potential are trustworthy and we shall discuss them below.

Also, the fact that the maximum of $LT$ decreases with the applied magnetic field implies that the imaginary part of the potential, computed for instance within the worldsheet fluctuation formalism \cite{Noronha:2009da,Finazzo:2013rqy}, may be enhanced by the magnetic field and this would affect the thermal width of heavy quarkonia in a strongly coupled plasma.

\begin{figure}[tbp]
\begin{center}
\begin{tabular}{cc}
\includegraphics[width=0.45\textwidth]{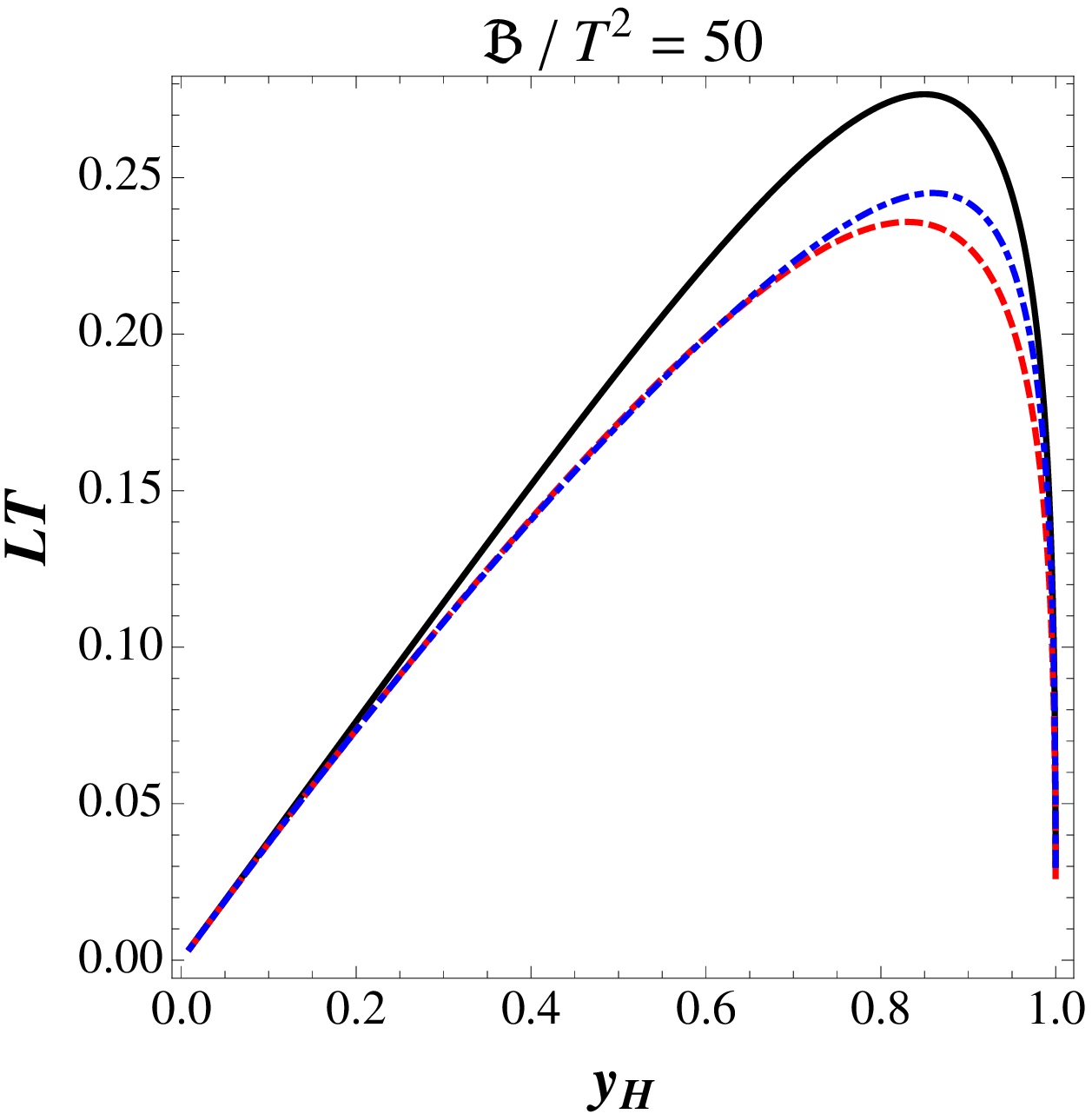} & %
\includegraphics[width=0.45\textwidth]{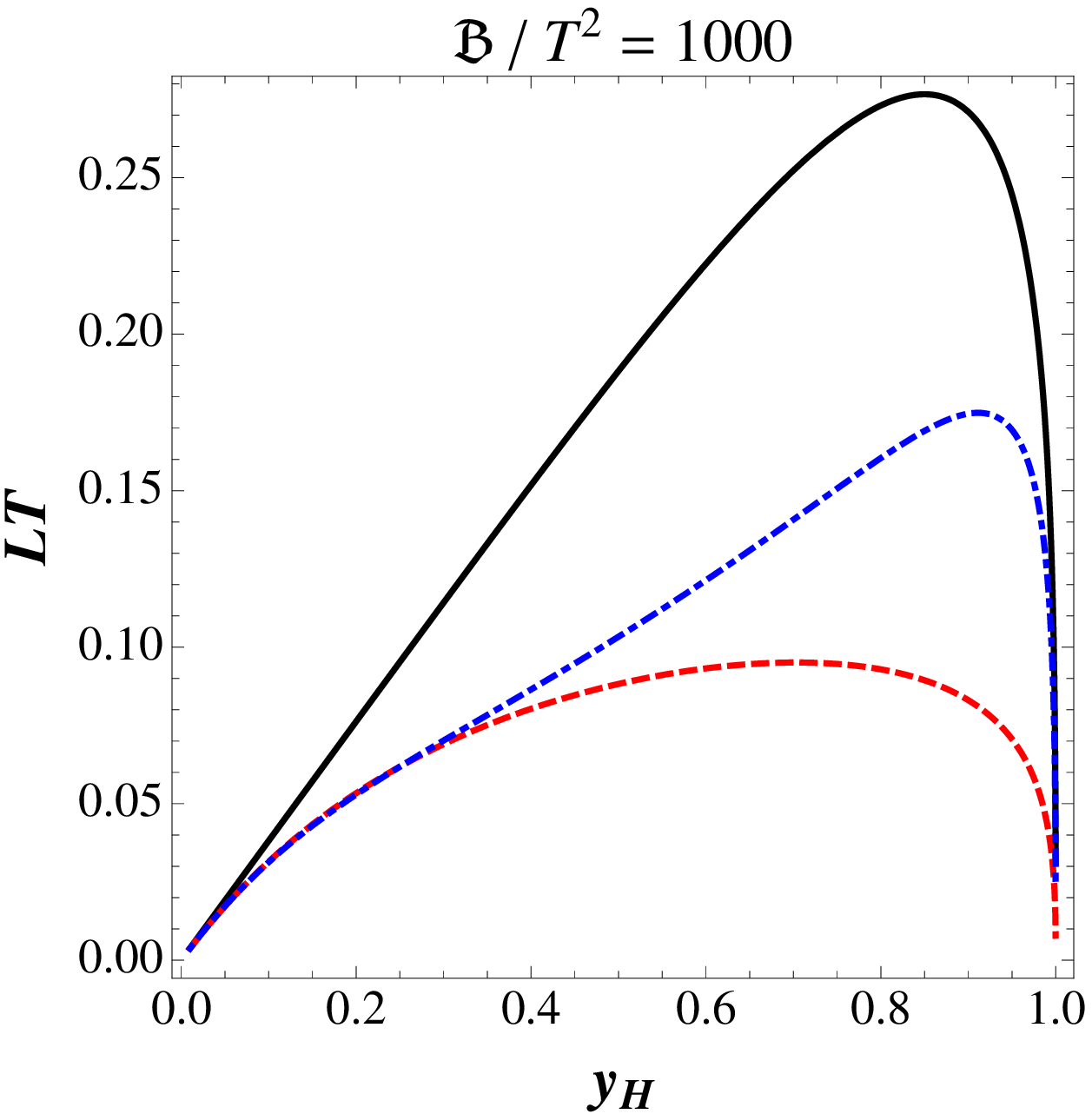} \\
&
\end{tabular}
\caption{(Color online) Interquark separation $LT$ versus the rescaled horizon $y_H$ (see Appendix \ref{apb}). In the left panel $\mathcal{B}/T^2=50$ while for the right panel $\mathcal{B}/T^2=1000$. For both panels the solid black line corresponds to the isotropic SYM case while the dashed red line (dotted-dashed blue line) corresponds to the case of anisotropic SYM with $Q\bar{Q}$ axis perpendicular (parallel) to the magnetic field axis.}
\label{fig4}
\end{center}
\end{figure}

\begin{figure}[tbp]
\begin{center}
\begin{tabular}{cc}
\includegraphics[width=0.45\textwidth]{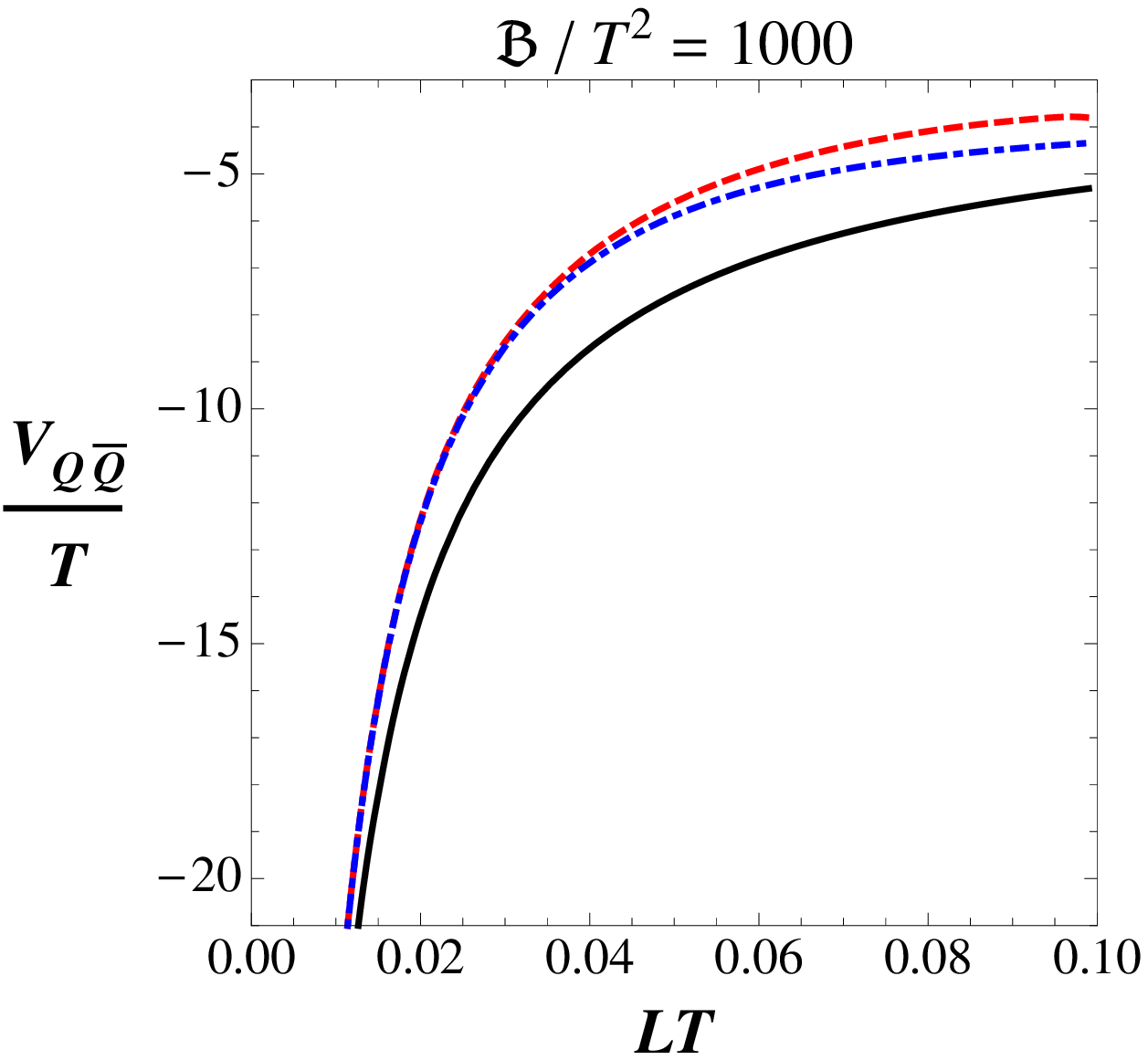} & %
\includegraphics[width=0.45\textwidth]{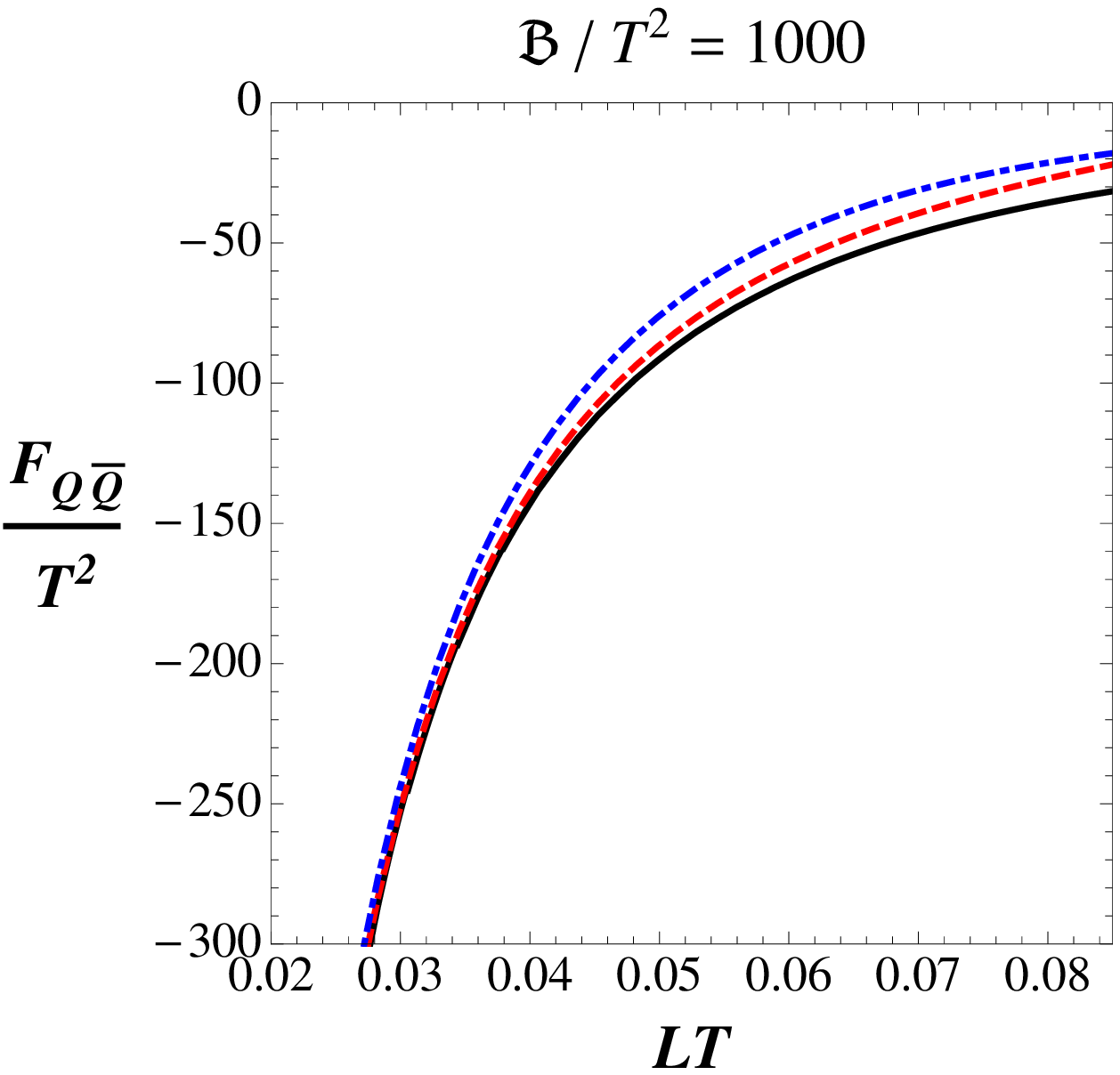} \\
&
\end{tabular}%
\caption{(Color online) Anisotropy induced by a strong magnetic field $\mathcal{B}/T^2=1000$ in the heavy quark potential (left panel) and the corresponding force (right panel) experience by a $Q\bar{Q}$ pair in a strongly-coupled $\mathcal{N}=4$ SYM plasma. The solid black lines correspond to the isotropic result $V_{Q\bar{Q}}^{\mathcal{B}=0}$ and isotropic force $F_{Q\bar{Q}}^{\mathcal{B}=0}=-dV_{Q\bar{Q}}^{\mathcal{B}=0}/dL$, the dashed red lines correspond to the perpendicular potential $V_{Q\bar{Q}}^\perp$ and perpendicular force $F_{Q\bar{Q}}^\perp=-dV_{Q\bar{Q}}^\perp/dL$, and the dotted-dashed curves correspond to the parallel potential $V_{Q\bar{Q}}^\parallel$ and force $F_{Q\bar{Q}}^\parallel=-dV_{Q\bar{Q}}^\parallel/dL$. In this plot $\alpha'=1$. }
\label{fig5}
\end{center}
\end{figure}

The combined effects from nonzero temperature and magnetic field on the heavy quark potential (left panel) and the corresponding force between the quarks (right panel) can be seen in Fig.\  \ref{fig5}. We found that the anisotropy in the heavy quark potential (and the force) induced by the magnetic field only becomes relevant for very large values of the field. In fact, in Fig.\ \ref{fig5} we have set $\mathcal{B}/T^2=1000$ to better illustrate the effects. The solid black lines correspond to the isotropic result for the potential $V_{Q\bar{Q}}^{\mathcal{B}=0}$ and its respective force, the dashed red lines correspond to the perpendicular potential $V_{Q\bar{Q}}^\perp$ and perpendicular force, and the dotted-dashed curves correspond to the parallel potential $V_{Q\bar{Q}}^\parallel$ and parallel force (in this plot $\alpha'=1$). By comparing Fig.\ \ref{fig5} and Fig.\ \ref{fig2} one can see that, roughly, the overall effect of the temperature is to shift the parallel and perpendicular potentials upwards with respect to the isotropic result. However, the pattern found at $T=0$ regarding the corresponding forces between the quarks is maintained, i.e., the force experienced by the quarks is the weakest when the pair axis is aligned with the magnetic field. Therefore, at least in the case of strongly coupled $\mathcal{N}=4$ SYM, we find that the inclusion of a magnetic field generally weakens the attraction between heavy quarks in the plasma.

\section{Conclusions and Outlook}
\label{conclusion}

In this paper we have studied how the inclusion of a constant magnetic field $\mathcal{B}$ affects the interaction between heavy $Q\bar{Q}$ pairs in strongly coupled $\mathcal{N}=4$ SYM theory both at zero and finite temperature by computing rectangular Wilson loops using the holographic correspondence. The magnetic field makes the heavy quark potential and the corresponding force anisotropic and we found that the attraction between the heavy quarks weakens in the presence of the magnetic field (both at $T=0$ and $T\neq 0$). Although, in practice, in the model considered here this effect only becomes relevant when $\mathcal{B}/T^2$ is extremely large \cite{Critelli:2014kra,Basar:2012gh}. 

We note that Ref.\ \cite{Chernicoff:2012bu} studied the anisotropy in the heavy quark potential induced by a nontrivial axion field in the bulk \cite{Mateos:2011tv} and found a reduction in the binding energy of the $Q\bar{Q}$ pair. This result is consistent with ours even though the source of anisotropy used in \cite{Chernicoff:2012bu} is different than the one used here (the constant magnetic field). This agreement between different anisotropic holographic models has also been found to hold in the case of transport coefficients since the shear viscosity coefficient along the direction of anisotropy computed in the axion-induced model \cite{Rebhan:2011vd} and in Ref.\ \cite{Critelli:2014kra} display the same qualitative behavior.

One may think that results in this paper give support to the idea that in a strongly coupled plasma deconfinement is facilitated by the inclusion of a magnetic field. However, such a conclusion may only be properly drawn in the case where the underlying gauge theory is not conformal at $T=0$ and $B=0$. In fact, the lattice results of Ref.\ \cite{Bonati:2014ksa} show that in QCD in a magnetic field at $T=0$ the absolute value of the Coulomb coupling in the direction of the magnetic field is enhanced with respect to its vacuum value while this coupling is suppressed in the case perpendicular to the magnetic field. On the other hand, the string tension perpendicular to the field is enhanced with respect to its vacuum value while the string tension parallel to the field is suppressed. This illustrates how complicated the effects of a magnetic field-induced anisotropy can be in a gauge theory with a mass gap.

It is interesting to see that the relative behavior we found between the perpendicular and parallel potentials in strongly coupled SYM qualitatively agrees with the perturbative calculation carried out in \cite{Miransky:2002rp} for the Coulomb-like part of the quark-antiquark potential in QCD at zero temperature, with the absolute value of the perpendicular potential being suppressed with respect to the potential in the direction of the magnetic field, which in turn is consistent with the behavior found on the lattice \cite{Bonati:2014ksa}. However, notice that SYM and QCD are very different theories in the vacuum and that, in particular, there is no confinement in SYM while in QCD for large quark-antiquark separations the potential becomes linear in the quark-antiquark separation (in the absence of dynamical quark flavors), instead of Coulomb-like. 

Moreover, even for the Coulomb-like part of the QCD quark-antiquark potential at zero temperature, the overall behavior is different than what we have found here for the strongly-coupled $\mathcal{N}=4$ SYM theory. In fact, in our case both the parallel and perpendicular potentials are enhanced with respect to the isotropic, zero magnetic field case. Also, both the parallel and perpendicular forces between the quark and the antiquark are suppressed with respect to the $\mathcal{B}=0$ case. In fact, if one rewrites the potential as $V=-\alpha(\sqrt{\mathcal{B}}L)/L=-(0.228+f(\sqrt{\mathcal{B}}L))/L$, one can see from Fig.\ \ref{fig2} that in the ultraviolet limit $L\rightarrow 0$ the potentials and forces go back to the vacuum result of Ref.\ \cite{maldacena}. However, for finite quark-antiquark separations, the effect of the magnetic field becomes relevant and the suppression observed for the parallel and perpendicular forces with respect to the isotropic, zero magnetic field case may be then related to some type of screening effect due to an effective change in the ``coupling constant" $\alpha(\sqrt{\mathcal{B}}L)$ due to the presence of the magnetic field.

It would be interesting to study modifications of the holographic setup addressed here and consider systems that are not conformal at $T=0$ and $\mathcal{B}=0$. For instance, consider a confining theory at zero temperature and finite magnetic field with confinement scale $\Lambda$. In this case, there is already a relevant dimensionless ratio $\mathcal{B}/\Lambda^2$ and, for instance, one can study how the mass gap of the theory is affected by the presence of the magnetic field and also how the area law of the rectangular Wilson loop becomes anisotropic and can be used to define a string tension for the heavy quark potential that depends on the angle between the $Q\bar{Q}$ pair and the magnetic field direction.

Such a model could be easily constructed following the bottom up studies in \cite{Gursoy:2007cb,Gursoy:2007er,Gursoy:2008bu,Gubser:2008ny,Noronha:2009ud} this time involving a dynamical metric, a scalar field, and a vector field in the bulk. The parallel and perpendicular potentials computed in this non-conformal model could be more easily compared to the lattice QCD study of Ref.\ \cite{Bonati:2014ksa}. Moreover, since such models are tuned to describe some of the thermodynamical properties of the strongly coupled QCD plasma found on the lattice \cite{Borsanyi:2010cj}, after the inclusion of the magnetic field, one could directly study in these models how the external field affects the deconfinement transition \cite{Bali:2011qj,Bali:2014kia} and also the role played by the paramagnetic behavior of QCD matter \cite{Fraga:2012ev} on the determination of other quantities, which also has been studied on the lattice \cite{Bonati:2013lca,Bonati:2013vba,Bali:2013owa}. Moreover, one could also investigate in such a model how the external field modifies other dynamical observables \cite{CasalderreySolana:2011us} that could be relevant to the phenomenology of the QGP formed in heavy ion collisions. We hope to address these questions in the near future.

\acknowledgments

We thank S.~I.~Finazzo, M.~Strickland, I.~A.~Shovkovy, and D.~Trancanelli for comments about the manuscript. This work was supported by Funda\c c\~ao de Amparo \`a Pesquisa do Estado de S\~ao Paulo (FAPESP) and Conselho Nacional de Desenvolvimento Cient\'ifico e Tecnol\'ogico (CNPq).

\appendix
\section{Coordinate transformations}
\label{apa}

In this Appendix we list the different coordinate systems used in this paper and how one may write the metric of $AdS_5$ spacetime in each one of them. A common way of expressing the $AdS_5$ metric in the context of the holographic correspondence is through the explicitly conformal coordinate system below
\begin{align}
ds^2=\frac{\ell^2}{U^2}(dU^2-dt^2+dx^2+dy^2+dz^2),
\label{a1}
\end{align}
where the boundary of the $AdS_5$ space is at $U=0$. Defining the coordinate transformation
\begin{align}
\bar{r}:=\frac{\ell^2}{U},
\label{a2}
\end{align}
one may rewrite the $AdS_5$ metric as follows
\begin{align}
ds^2=\frac{\ell^2}{\bar{r}^2}d\bar{r}^2+\frac{\bar{r}^2}{\ell^2}(-dt^2+dx^2+dy^2+dz^2),
\label{a3}
\end{align}
where the boundary of the $AdS_5$ space is now at $\bar{r}\rightarrow\infty$. This coordinate system is the one used in \cite{DK1} and in Sec.\ \ref{sec3} to obtain the finite temperature solutions.

Also, through the coordinate transformation
\begin{align}
r:=\frac{\bar{r}^2}{2\ell}=\frac{\ell^3}{2U^2},
\label{a4}
\end{align}
one can write the $AdS_5$ metric as
\begin{align}
ds^2=\frac{\ell^2}{4r^2}dr^2+\frac{2r}{\ell}(-dt^2+dx^2+dy^2+dz^2),
\label{a5}
\end{align}
where the boundary is at $r\rightarrow\infty$. We can further define the light-cone coordinates
\begin{align}
u:=\frac{z+t}{\sqrt{2}},\,\,\,v:=\frac{z-t}{\sqrt{2}},
\label{a6}
\end{align}
in terms of which (\ref{a5}) is rewritten as follows
\begin{align}
ds^2=\frac{\ell^2}{4r^2}dr^2+\frac{4r}{\ell}dudv+\frac{2r}{\ell}(dx^2+dy^2)\,.
\label{a7}
\end{align}
This coordinate system is the one used in \cite{DK2} and in Sec.\ \ref{sec2} to study the zero temperature solution of the model.

\section{Wilson loops in $\mathcal{N}=4$ SYM}
\label{apb}

For the sake of completeness, in this Appendix we give a brief review of the holographic computation of rectangular Wilson loops in SYM at finite temperature \cite{sonne,rey} without magnetic fields. We shall closely follow the discussions in Section 5.1 of Ref.\ \cite{jorge}. At finite $T$ and $B=0$, the background giving an holographic description of thermal SYM is the $AdS_5$-Schwarzschild metric
\begin{align}
ds^2=\frac{\ell^2}{\bar{r}^2f(\bar{r})}d\bar{r}^2-\frac{\bar{r}^2f(\bar{r})}{\ell^2}d\bar{t}^2+ \frac{\bar{r}^2}{\ell^2}(d\bar{x}^2+d\bar{y}^2+d\bar{z}^2), \,\,\,f(\bar{r})=1-\frac{\bar{r}_H^4}{\bar{r}^4},
\label{b1}
\end{align}
where the boundary is at $\bar{r}\rightarrow\infty$ and the horizon is at $\bar{r}=\bar{r}_H$. Rescaling $\bar{r}=:4\bar{r}_H(r-3/4)$, $(\bar{t},\bar{x},\bar{y},\bar{z})=:(t,x,y,z)/4\bar{r}_H$ and adopting units where $\ell=1$, one rewrites (\ref{b1}) as follows
\begin{align}
ds^2=\frac{dr^2}{\left(r-\frac{3}{4}\right)^2f(r)}-\left(r-\frac{3}{4}\right)^2f(r)dt^2+\left(r-\frac{3}{4}\right)^2 d\vec{x}^{\,2}, \,\,\,f(r)=1-\frac{1}{\left[4\left(r-\frac{3}{4}\right)\right]^4},
\label{b2}
\end{align}
where the boundary is at $r\rightarrow\infty$ and the horizon is now at $r=1$. From (\ref{b2}), the Hawking temperature reads
\begin{align}
T=\frac{\sqrt{-g'_{tt}\,g^{rr}\,'}}{4\pi}\biggr|_{r=1}=\frac{1}{4\pi},
\label{b3}
\end{align}
which is the same constant temperature obtained before in Eq.\ (\ref{3.13}) for the magnetic backgrounds. Indeed, one can check numerically that the magnetic backgrounds at finite temperature derived in Sec.\ \ref{sec3} converge for the metric (\ref{b2}) in the limit of zero magnetic field, as it should be.

The formal expressions for the interquark distance and the heavy quark potential (we set $\alpha'=1$ below) as functions of the parameter $r_0$ are given by\footnote{For instance, one may obtain these expressions by replacing $h(r)\rightarrow (r-3/4)^2f(r)$ and $e^{2W(r)}\rightarrow (r-3/4)^2$ in Eqs.\ (\ref{3.22}) and (\ref{3.23}).}
\begin{align}
L_{Q\bar{Q}}^{(T\neq 0)}(r_0)&= 32\sqrt{(4r_0-3)^4-1}\int_{r_0}^{\infty} \frac{dr}{\sqrt{[(4r-3)^4-1][(4r-3)^4-(4r_0-3)^4]}},\label{b4}\\
V_{Q\bar{Q}}^{(T\neq 0)}(r_0)&= \frac{1}{\pi}\left[ \int_{r_0}^\infty dr\left(\sqrt{\frac{(4r-3)^4-1}{(4r-3)^4-(4r_0-3)^4}}-1\right) - \int_0^{r_0}dr\right].\label{b5}
\end{align}
Defining the new integration variable $R:=4r-3$ and also the constant $R_0:=4r_0-3$, we rewrite (\ref{b4}) and (\ref{b5}) as follows
\begin{align}
L_{Q\bar{Q}}^{(T\neq 0)}(R_0)&= 8\sqrt{R_0^4-1}\int_{R_0}^{\infty} \frac{dR}{\sqrt{(R^4-1)(R^4-R_0^4)}},\label{b6}\\
V_{Q\bar{Q}}^{(T\neq 0)}(R_0)&= \frac{1}{4\pi}\left[ \int_{R_0}^\infty dR\left(\sqrt{\frac{R^4-1}{R^4-R_0^4}}-1\right) -R_0-3\right]\,.\label{b7}
\end{align}
Defining now $y:=R/R_0$ and also $y_H:=1/R_0$, one rewrites (\ref{b6}) and (\ref{b7}) as follows
\begin{align}
L_{Q\bar{Q}}^{(T\neq 0)}(y_H)&= 8y_H\sqrt{1-y_H^4}\int_{1}^{\infty} \frac{dy}{\sqrt{(y^4-y_H^4)(y^4-1)}},\label{b8}\\
V_{Q\bar{Q}}^{(T\neq 0)}(y_H)&= \frac{1}{4\pi y_H}\left[ \int_{1}^\infty dy\left(\sqrt{\frac{y^4-y_H^4}{y^4-1}}-1\right) - 1 - 3y_H\right]\,.\label{b9}
\end{align}
Let us denote the integrals in (\ref{b8}) and (\ref{b9}) by $\mathcal{I}_1$ and $\mathcal{I}_2$, respectively. In what follows, we are going to express these integrals in terms of Gaussian hypergeometric functions, by using the following integral representation
\begin{align}
_2F_1(a,b;c;z)=\frac{\Gamma(c)}{\Gamma(b)\Gamma(c-b)}\int_0^1 dx\, x^{b-1}\, (1-x)^{c-b-1}\, (1-zx)^{-a},
\label{b10}
\end{align}
which is valid for $\textrm{Re}[c]>\textrm{Re}[b]>0$ and $|z|<1$.

Defining the new integration variable $x:=y^{-4}$, one obtains for the integral in (\ref{b8})
\begin{align}
\mathcal{I}_1=\int_{1}^{\infty} \frac{dy}{\sqrt{(y^4-y_H^4)(y^4-1)}}&= \frac{1}{4}\int_0^1 dx\, x^{-1/4}\, (1-x)^{-1/2}\, (1-xy_H^4)^{-1/2}\nonumber\\
&=\frac{1}{4}\frac{\Gamma(3/4)\Gamma(1/2)}{\Gamma(5/4)}\,_2F_1\left(\frac{1}{2},\frac{3}{4};\frac{5}{4};y_H^4\right)\nonumber\\
&\approx 0.599\,_2F_1\left(\frac{1}{2},\frac{3}{4};\frac{5}{4};y_H^4\right).
\label{b11}
\end{align}
For the integral in (\ref{b9}),
\begin{align}
\mathcal{I}_2=\int_{1}^\infty dy\left(\sqrt{\frac{y^4-y_H^4}{y^4-1}}-1\right) = \frac{1}{4}\int_0^1 dx\, x^{-5/4} \left[ (1-x)^{-1/2}\, (1-xy_H^4)^{1/2}-1\right],
\label{b12}
\end{align}
we employ the following regularization scheme\footnote{Note this regularization procedure involves commuting the limit $\lambda\rightarrow 0$ with the integral.} in order to allow the use of the integral representation (\ref{b10})
\begin{align}
\mathcal{I}_2^{\textrm{reg}}&=\frac{1}{4}\lim_{\lambda\rightarrow 0}\int_0^1 dx\, x^{-5/4+\lambda} \left[ (1-x)^{-1/2}\, (1-xy_H^4)^{1/2}-1\right]\nonumber\\
&=\frac{1}{4}\lim_{\lambda\rightarrow 0}\left[ \frac{\Gamma(-1/4+\lambda)\Gamma(1/2)}{\Gamma(1/4+\lambda)}\, _2F_1(-\frac{1}{2},-\frac{1}{4}+\lambda;\frac{1}{4}+\lambda;y_H^4) + \frac{4}{1-4\lambda} \right]\nonumber\\
&\approx -0.599\,_2F_1\left(-\frac{1}{2},-\frac{1}{4};\frac{1}{4};y_H^4\right)+1\,.
\label{b13}
\end{align}

Substituting (\ref{b11}) into (\ref{b8}) and (\ref{b13}) into (\ref{b9}), one obtains, respectively
\begin{align}
L_{Q\bar{Q}}^{(T\neq 0)}(y_H)&\approx 4.792\, y_H\sqrt{1-y_H^4}\,_2F_1\left(\frac{1}{2},\frac{3}{4};\frac{5}{4};y_H^4\right),\label{b14}\\
V_{Q\bar{Q}}^{(T\neq 0)}(y_H)&\approx -\frac{0.048}{y_H}\,_2F_1\left(-\frac{1}{2},-\frac{1}{4};\frac{1}{4};y_H^4\right)- \frac{3}{4\pi},\label{b15}
\end{align}
with $|y_H^4|<1$. Eqs.\ (\ref{b14}) and (\ref{b15}) were employed to obtain numerically the parametric SYM curve in Fig.\ \ref{fig5}.

As a final remark, we mention that the values of $y_H$ considered in the parametric plots shown in Fig.\ \ref{fig5} were restricted to values below $y_H^{\textrm{max}}$, which is the value of $y_H$ where $LT$ reaches its maximum value.

\end{document}